\documentclass[10pt,aps,prd,twocolumn,notitlepage,nofootinbib,floatfix,
superscriptaddress]{revtex4-1}

\usepackage{amsmath, amsthm, amssymb, amsfonts, amsbsy, mathrsfs}
\usepackage{calligra,bm}
\usepackage{stmaryrd}

\usepackage[hidelinks,bookmarks=true]{hyperref}
\hypersetup{pdfstartview=FitH,pdfhighlight=/O,colorlinks=false}

\begin{document}

\title{Note on gravity at the boundary of an AdS vacuum}

\author{Justin C. Feng}
\affiliation{Centro de Astrofísica e Gravitação - CENTRA,
Departamento de Física, Instituto Superior Técnico - IST,
Universidade de Lisboa - UL,
Av. Rovisco Pais 1, 1049-001 Lisboa, Portugal}

%=======================================================================
%
%        ABSTRACT
%
%=======================================================================
\begin{abstract}
    In this note, I describe an attempt to construct a phenomenological gravitational model at the boundary of the AdS manifold from the variation of boundary terms in the gravitational action. I find that for an AdS vacuum in the bulk, geometric constraints require that the energy-momentum tensor has constant trace.
\end{abstract}

%=======================================================================

\maketitle

This brief note describes an attempt to construct a phenomenological model for classical gravity motivated by a variational principle for boundary terms in vacuum AdS spacetimes. The original hope in constructing this model was to obtain a model resembling those of braneworld models (see the reviews \cite{Maartens2010,*Brax2004,*Langlois2004,*Rubakov2001} and references contained therein) and \cite{Mukohyama2001,*Khoury2006} on the boundary of AdS spacetime. While such a model might be viable if bulk matter is included, this note demonstrates that if the bulk is assumed to be an AdS vacuum, geometric constraints require that trace of the boundary energy-momentum tensor is constant, limiting the phenomenological utility of such a model.

On manifolds with boundary, variational principles typically require boundary conditions for the degrees of freedom involved. In general relativity (GR), the following action in $N$ dimensions:
\begin{equation} \label{GravitationalActionN}
S_{G} = \frac{1}{2 \kappa}\int_{U} d^N x \sqrt{-\mathfrak{g}} \,(\mathfrak{R} - 2 \Lambda) + \frac{\varepsilon}{ \kappa} \int_{\partial U} d^Dy \sqrt{-g} \, K
\end{equation}

\noindent yields the vacuum Einstein equation $\mathfrak{G}_{\mu \nu} = - \Lambda g_{\mu \nu}$ with cosmological constant $\Lambda$ when the induced metric $g_{ij}$ is held fixed at the boundary \cite{GibbonsHawking1977,York1972}. Here $D:=N-1$, $K$ is the mean curvature of the boundary $\partial U$, and $\varepsilon=+1$ for timelike $\partial U$. Points in $U$ and $\partial U$ are respectively denoted $x$ and $y$. For convenience, define $D_1:=D-1$ and $D_2:=D-2$.

For the anti de Sitter (AdS) manifold, defined here as a Lorentzian manifold with the topology of AdS spacetime (the properties of which are described in \cite{HawkingEllis}), one can add a boundary term which does not require boundary conditions on the timelike conformal boundary at infinity. If the induced metric on the boundary is dynamical, one can in principle add geometric boundary terms yielding equations of motion on the boundary that determine (at least partly) the boundary conditions. Such ``mixed'' boundary conditions have been investigated before in the context of holographic renormalization in the AdS/CFT correspondence---see for instance \cite{Compere2008,Siopsis2013}. Here, I consider a purely classical problem, focusing primarily on the equations of motion generated by boundary terms up to linear order in the Ricci scalar.

Consider a boundary term of the following form:
\begin{equation} \label{GravitationalActionBdy}
\sigma = - \frac{\varepsilon}{ \kappa} \int_{\partial U} d^Dy \sqrt{-g} \left[D_1 \, k + {R}/2\bar{\kappa}\right] ,
\end{equation}

\noindent where $k D$ is the mean curvature for the conformal boundary of AdS spacetime, $\bar{\kappa}$ is a constant parameter, and $\bar{R}$ is the Ricci scalar of the boundary $\partial U$ (barred quantities are defined with respect to the boundary $\partial U$). Except for the parameter $\bar{\kappa}$, which is left arbitrary, Eq. \eqref{GravitationalActionBdy} has the form of the holographic counterterm up to linear order in the boundary Ricci curvature \cite{Balasubramanian1999,*Emparan1999,*deHaro2000,*Taylor2000} (to recover the holographic counterterm, set $\bar{\kappa}= \sqrt{2 |\Lambda| (D_2){^2} /(D \, D_1)}$). The full variation of the action $S_{G}^\prime:=S_{G}+\sigma$ takes the form
\begin{equation} \label{VarGravitationalActionN2}
\begin{aligned}
\delta S_{G}^\prime = & \frac{1}{2 \kappa}\int_{U} d^N x \sqrt{-\mathfrak{g}} \, (\mathfrak{G}_{\mu \nu} + \Lambda \, g_{\mu \nu}) \delta g^{\mu \nu}\\
& + \frac{\varepsilon}{ \kappa} \int_{\partial U} d^Dy \sqrt{-g}\left[\tau_{ij}/2 - {G}_{ij}/2\bar{\kappa}\right] \delta g^{ij} ,
\end{aligned}
\end{equation}

\noindent where ${G}_{ij}$ is the Einstein tensor on $\partial U$ (lowercase Latin indices refer to the coordinate basis on the boundary) and the bulk induced energy-momentum tensor is defined:
\begin{equation} \label{EffectiveEMT}
\tau_{ij} := K_{ij} - [K - D_1 k] g_{ij}.
\end{equation}

\noindent One can find a coordinate basis on the conformal boundary of pure AdS spacetime in which the extrinsic curvature tensor $K{^i}{_j}$ is diagonal and the diagonal elements all have the same value $k := \sqrt{2 |\Lambda|/(D \, D_1)}$; it is straightforward to check that $\tau_{ij}=0$ on the boundary of AdS spacetime. This property generalizes: the extrinsic curvature $K_{ij}$ (and consequently $\tau_{ij}$) of the conformal boundary at infinity for an AdS vacuum is proportional to $g_{ij}$ (See prop. 2.8 in \cite{Lebrun1982}).

Demanding $\delta S_{G}^\prime = 0$ under variations of the metric,
\begin{eqnarray}
\mathfrak{G}_{\mu \nu} &= - \Lambda \, \mathfrak{g}_{\mu \nu} \label{EOMBulk}, \\
{G}_{ij} &=  \bar{\kappa} \, \tau_{ij}, \label{EOMBoundary}
\end{eqnarray}

\noindent where Eq. \eqref{EOMBoundary} is evaluated on the conformal boundary. One can verify that the AdS spacetime is indeed a solution for Eqs.~\eqref{EOMBulk} and \eqref{EOMBoundary}. The action $S_{G}^\prime$ therefore supplies a set of boundary conditions for the metric tensor on the conformal boundary of the AdS manifold; at the boundary, the induced metric and extrinsic curvature must satisfy a condition in the form of an Einstein equation on the boundary. The induced energy-momentum tensor $\tau_{ij}$ is subject to a geometric ``momentum'' constraint (using the language of the 3+1 formalism) ${\nabla}_i(K^{ij}-K \gamma^{ij})=0$ and a ``Hamiltonian'' constraint \cite{Bautier2000,Fischetti2012} which comes from the bulk equation of motion $\mathfrak{G}_{\mu \nu} n^\mu n^\nu = - \Lambda$ ($n^\mu$ being the unit normal vector to $\partial U$). The ``Hamiltonian'' constraint may be written in terms of $K_{ij}$ and the Ricci scalar $R$ on $\partial U$:
\begin{equation} \label{HamiltonianConstraint}
2 \Lambda = {R} + K{^i}{_j} \, K{^j}{_i} - K^2.
\end{equation}

\noindent The original hope in this construction was that one could construct a nontrivial matter model for $\tau_{ij}$, the dynamics being determined by the ``Hamiltonian'' and ``momentum'' constraints (for instance, a perfect fluid with equation of state determined by Eq. \eqref{HamiltonianConstraint}). However, if the bulk geometry is vacuum, $\tau_{ij} = \left({\tau} / D \right) \, g_{ij}$ at the conformal boundary for some scalar $\tau$; the ``momentum'' constraint and contracted Bianchi identity imply that $\tau$ must be a constant, the value of which is determined by the Hamiltonian constraint.

The discussion so far is only formal, since the components of the metric $g_{\mu \nu}$ and induced metric $g_{ij}$ diverge at the boundary; this can be seen in Fefferman-Graham coordinates \cite{FeffermanGraham} near the boundary, in which the line element near $\partial U$ for a bulk geometry satisfying \eqref{EOMBulk} takes the form:
\begin{equation} \label{LEGF}
ds^2 = \frac{l^2}{Z^2}\left[dZ^2 + \left(\tilde{g}_{ij}+ \mathcal{O}(Z)\right) \, dx^i \, dx^j \right],
\end{equation}

\noindent where the coordinate $Z=0$ at the boundary $\partial U$, $l^2=-d(d-1)/2\Lambda$, and $\tilde{g}_{ij}=z^2 \, g_{ij}$ (defining $z:=Z/l$) is a rescaled (and finite) induced metric at $\partial U$. It is straightforward to show that under the rescaling $\tilde{g}_{ij}=z^2 \, g_{ij}$, the rescaled Ricci tensor $\tilde{R}_{ij}$ satisfies $\tilde{R}_{ij}={R}_{ij}$,
but the Ricci scalar $\tilde{R}$ satisfies ${R} = z^2 \, \tilde{R}$; note that $g_{ij} \, {R}  = \tilde{g}_{ij} \, \tilde{R}$, so $\tilde{G}_{ij}={G}_{ij}$. It follows that Eq. \eqref{EOMBoundary} implies $\tilde{\tau}_{ij}={\tau}_{ij}$,
so that $\tilde{\tau}_{ij} = \left(\tilde{\tau} / D \right) \, \tilde{g}_{ij}$ under this rescaling. Rescaling of Eq. \eqref{EOMBoundary} yields $\tilde{G}_{ij} = \bar{\kappa} \, \tilde{\tau}_{ij}$, and one finds that $\tilde{\tau}$ must also be a constant.

I now consider what happens when boundary matter is inserted by hand. One might imagine adding a boundary term $\sigma_m=\sigma_m[\varphi,g^{\cdot \cdot}]$ which serves as an action for the matter degrees of freedom $\varphi=\varphi(y)$ on the boundary, but for the argument presented here, it suffices to simply add an energy-momentum tensor (the bulk is assumed to be an AdS vacuum):
\begin{equation} \label{EOMBoundaryMatterRS}
\tilde{G}_{ij} =  \bar{\kappa} \left( \tilde{\tau}_{ij} + \tilde{T}_{ij} \right).
\end{equation}

\noindent For generality, I do not assume the vanishing of $\tilde{T} := \tilde{g}^{ij} \, \tilde{T}_{ij}$. To obtain the ``Hamiltonian'' constraint under the rescaling $\tilde{g}_{ij}=z^2 \, g_{ij}$, one begins by inverting Eq. \eqref{EffectiveEMT} to obtain $K_{ij} = \tau_{ij} + [k - \tau / D_1] \gamma_{ij}$.
The trace of Eq. \eqref{EOMBoundaryMatterRS} yields ${R} = -2z^2 \bar{\kappa}(\tilde{\tau} + \tilde{T})/D_2$. In terms of rescaled quantities, Eq. \eqref{HamiltonianConstraint} may be rewritten as the following constraint for the rescaled energy-momentum tensor $\tilde{\tau}_{ij}$ (note ${\tau}{^i}{_j} = z^2 \, \tilde{\tau}{^i}{_j}$):
\begin{equation} \label{EMTConstraint}
z^2 \, \tilde{\tau}{^i}{_j} \, \tilde{\tau}{^j}{_i} = \tilde{\tau} \left(z^2 \tilde{\tau}/{D_1} - 2 k\right) + {2 \bar{\kappa} (\tilde{\tau} + \tilde{T})}/{D_2} .
\end{equation}

\noindent In the $z \rightarrow 0$ limit, this yields the following constraint on the trace at the boundary $\partial U$:
\begin{equation} \label{EMTConstraintBdy}
 \left(D_2 k - \bar{\kappa} \right) \tilde{\tau} = \bar{\kappa} \tilde{T}.
\end{equation}

\noindent It follows that for $\tilde{T} = 0$, one has $\tilde{\tau} = 0$ and $\tilde{\tau}_{ij} = 0$. If in addition one has $\tilde{T}_{ij} = 0$, then one recovers the vacuum Einstein equations on the boundary. If $\tilde{T} \neq 0$, the constraint \eqref{EMTConstraintBdy} indicates that $\tilde{\tau} \propto \tilde{T}$.
However, the ``momentum'' constraint $\tilde{\nabla}^i \tilde{\tau}_{ij} = 0$ for a bulk AdS vacuum implies that $\tau$ must be a constant and it follows that $\tilde{T}$ must also be constant. It seems that the properties of the boundary for an AdS vacuum severely restrict the boundary matter models compatible with the gravitational theory described in this note; the phenomenological utility of the boundary theory is limited.

%=======================================================================
%        ACKNOWLEDGMENTS
%=======================================================================

         \begin{acknowledgments}
         I thank
         Vitor Cardoso,
         Sante Carloni,
         David Hilditch,
         Richard Matzner,
         Masato Minamitsuji,
         Shinji Mukohyama,
         and
         Christiana Pantelidou
         for feedback and reference suggestions. I also thank an anonymous referee for making me aware of ref. \cite{Lebrun1982}. This work was supported by FCT Grant Nos. PTDC/MAT-APL/30043/2017 and UIDB/00099/2020.
         \end{acknowledgments}

%=======================================================================
%        BIBLIOGRAPHY
%=======================================================================

\bibliography{GBAdS}

%=======================================================================

\end{document}